\documentclass[12pt,preprint]{aastex}
\shortauthors{JENKINS \& TRIPP.}
\shorttitle{NI II F-VALUES}
\begin{document}
\title{Measurements of the $f$-Values of the Resonance Transitions of
Ni~II at 1317.217\AA\ and 1370.132\AA\footnote{Based on observations
from the NASA/ESA {\it Hubble Space Telescope\/} obtained at the Space
Telescope Science Institute, which is operated by the Association of
Universities for Research in Astronomy, Inc., under NASA contract
NAS5-26555}}
\author{Edward B. Jenkins}
\affil{Princeton University Observatory,
Princeton, NJ 08544-1001}
\email{ebj@astro.princeton.edu}
\author{Todd M. Tripp}
\affil{Department of Astronomy, University of Massachusetts\\
Amherst, MA 01003-9305}
\email{tripp@fcrao1.astro.umass.edu}

\begin{abstract}
We have retrieved high-resolution UV spectra of 69 hot stars from the
{\it HST\/} archive and determined the strengths of the interstellar
Ni~II absorption features at 1317.217$\,$\AA\ arising from the ground
$3d^9~^2D_{5/2}$ electronic state to the $3d^8(^1G)4p~^2F^{\rm
\,o}_{5/2}$ excited level.  We then compared them to absorptions to
either the $3d^8(^3F)4p~^2D^{\rm \,o}_{5/2}$ or $3d^8(^3P)4p~^2P^{\rm
\,o}_{3/2}$ upper levels occurring at, respectively, $\lambda =
1741.553\,$\AA\ (covered in the spectra of 21 of the stars) and
$1370.132\,$\AA\ (seen for the remaining 48 stars).  All spectra were
recorded by the either the E140M, E140H, or E230H gratings of the Space
Telescope Imaging Spectrograph.  By comparing the strengths of the two
lines in each spectrum and evaluating a weighted average of all such
comparisons, we have found that the $f$-value of the 1317$\,$\AA\ line
is $1.34\pm 0.019$ times the one at 1741$\,$\AA\ and $0.971\pm 0.014$
times that of the one at 1370$\,$\AA.  We adopt as a comparison standard
an experimentally determined $f$-value for the 1741$\,$\AA\ line (known
to 10\% accuracy), so that $f(1317\,$\AA)~=~0.0571$\pm 0.006$.  It
follows from this $f$-value and our measured line-strength ratios that
$f(1370\,$\AA)~=~0.0588$\pm 0.006$.  As an exercise to validate our
methodology, we compared the 1317$\,$\AA\ transition to another Ni~II
line at 1454.842$\,$\AA\ to the $3d^8(^1D)4p~^2D^{\rm \,o}_{5/2}$ level
and arrived at an $f$-value for the latter that is consistent with a
previously measured experimental value to within the expected error.
\end{abstract}

\section{Introduction}\label{intro}

For investigations of the interstellar gases in either our Galaxy or
distant systems in the universe, nickel belongs to a class of elements
that is usually strongly depleted onto dust grains  (Savage \& Sembach
1996; Jenkins 2004), but also represents the important iron-peak group
that is synthesized mostly by Type~Ia supernovae at later stages in a
galaxy's chemical evolution  (Wheeler, Sneden, \& Truran 1989; McWilliam
1997).  For neutral regions having $N({\rm H~I})\gtrsim 10^{19.5}{\rm
cm}^{-2}$, most of the Ni atoms should be in the singly ionized form. 
Numerous Ni~II lines with differing strengths in the wavelength interval
$1300 < \lambda < 1900\,$\AA\ are well suited for deriving column
densities of singly-ionized Ni using a curve of growth.  Other Fe-group
elements such as Cr, Mn, Fe, Co and Zn also exhibit two or more
moderately strong lines for their most abundant, singly-ionized forms,
but most of the useful features are at $\lambda > 2000\,$\AA\ in the
rest frame, a wavelength range that might not be covered by some
observations.  (A few lines of Cu~II at shorter wavelengths are usually
too weak to observe.)

In the recent compilation by Morton  (2003) of UV and visible
transitions out of the ground electronic states of various atoms,
experimentally determined $f$-values are listed for 14 different
transitions of Ni~II, with numerical values that range from 0.001 to
0.08,\footnote{Many more $f$-values for Ni~II were listed in Morton's 
(1991) earlier survey of resonance lines, but these values were based on
the approximate theoretical $f$-values of Kurucz (communicated privately
to Morton).  In his publication of 2003, Morton chose not to list these
$f$-values because they had a much lower accuracy than the experimental
values.} based on the laser-induced fluorescence lifetime measurements
of Fedchak \& Lawler  (1999) for a few of them, with extensions to the
remaining ones using relative absorption strengths measured in the
laboratory by Fedchak, Wiese \& Lawler  (2000)  and supplemented by
Hubble Space Telescope ({\it HST\/}) recordings of absorption lines
measured by Zsarg\'o \& Federman  (1998).\footnote{Zsarg\'o \& Federman 
(1998) measured relative $f$-values and scaled them according the old
values for other Ni~II lines given in Morton  (1991).  Later, Fedchak \&
Lawler  (1999) recommended that those $f$-values should be scaled
downward by a factor 0.534, on the basis of changes in the $f$-values
indicated for comparison transitions at longer wavelengths.  Morton's 
(2003) compilation has adopted the $f$-values of Zsarg\'o \& Federman 
(1998) with this reduction factor.}  These measurements have made some
important improvements in the $f$-values of different Ni~II transitions
and have resolved some earlier apparent anomalies in the abundance
ratios of Ni to Fe in various distant gas systems  (Howk, Savage, \&
Fabian 1999).  Finally, theoretical calculations of the transition
probabilities for the $3d^9 - 3d^84p$ and $3d^84s - 3d^84p$ transition
arrays of Ni~II have been carried out by Fritzsche, Dong \& Gaigalas 
(2000).

One transition that has not yet been investigated experimentally is the
$3d^9~^2D_{5/2}\rightarrow 3d^8(^1G)4p~^2F^{\rm \,o}_{5/2}$ line of
Ni~II at 1317.217$\,$\AA.  As we show later, this line has a strength
about equal to those of the strongest lines of Ni~II at 1370.132$\,$\AA\
and 1741.553$\,$\AA\ to the upper level configurations
$3d^8(^3P)4p~^2P^{\rm \,o}_{3/2}$ and $3d^8(^3F)4p~^2D^{\rm \,o}_{5/2}$. 
 For extragalactic systems that have only enough Ni~II to allow a
detection using either the 1370$\,$\AA\ line or the one at 1741$\,$\AA,
the usefulness of these lines might be compromised in some circumstances
by either restrictions in the wavelength coverage or possible
interference from random lines from the Ly$\alpha$ forest when $z_{\rm
abs} \ll z_{\rm em}$.  Thus, having a redundant feature with about the
same strength at a different wavelength provides a useful alternative to
help overcome these limitations.

The strength of the transition at 1370$\,$\AA\ has not been measured in
the laboratory, but instead Zsarg\'o \& Federman  (1998) have compared
interstellar features for this line to those seen at longer wavelengths
for which laboratory data now exist [their original study made use of
the less accurate $f$-values given in Morton  (1991)].  This comparison,
however, was performed for only 2 velocity components appearing in the
spectrum of a single star ($\zeta$~Oph), and the interpretations of
these mildly saturated lines made use of curves of growth for velocity
parameters $b$ derived from the lines of other species  (Savage,
Cardelli, \& Sofia 1992).  Thus, while the $f$-value of the 1370$\,$\AA\
line has been determined astrophysically, it is reasonable to expect an
improvement in accuracy if more interstellar line data can be obtained.

Our objective of this study is to gather high-resolution spectra of hot
stars from the {\it Hubble Space Telescope\/} ({\it HST\/}) archive and
intercompare the strengths of the interstellar absorption features from
Ni~II at the wavelengths 1317, 1370, and 1741$\,$\AA.  The line at
1741$\,$\AA\ will be used as our fundamental comparison standard, since
its $f$-value has been measured in the laboratory  (Fedchak \& Lawler
1999) and thus is relatively trustworthy.  An important feature of our
comparisons is that the near similarity of line strengths for any given
target star makes our interpretations relatively immune to poorly
understood saturation effects, since such effects, if they exist, will
have roughly equal strengths for all three lines and thus will not
affect the ratios that we calculate.  The spectra that we use contain
either the line pair 1317 and 1370$\,$\AA, or the pair 1317 and
1741$\,$\AA.  Thus, we are not able to directly compare the 1370$\,$\AA\
line to the one at 1741$\,$\AA, but instead we must link it to the line
at 1317$\,$\AA.

A similar determination of the relative Ni~II $f$-values was carried out
by Ellison et al.  (2001) using a spectrum of a damped Ly$\alpha$ system
in front of a quasar.  We expect that our comparison should be more
accurate, since we are using many spectra and ones of high S/N.  Also,
in many cases our spectral resolution is much higher, which makes it
much less likely that we could be misled by the effects of narrow,
saturated velocity components.

\section{Data}\label{data}

\begin{figure}
\epsscale{.75}
\plotone{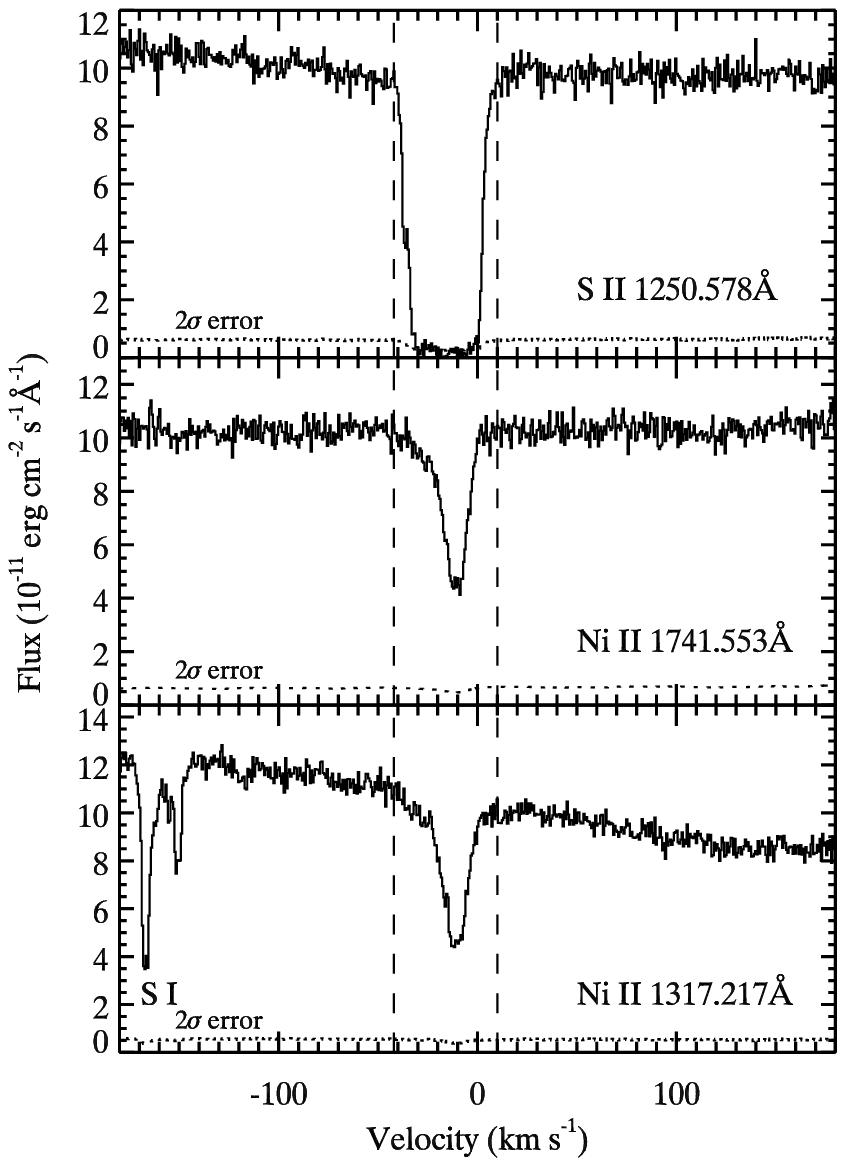}
\caption{One example of the absorption profiles for S~II (top panel) and
Ni~II (middle and lower panels), taken from the spectrum recorded for
the star HD~210839.  This spectrum was selected to illustrate the
importance of using the extent of the S~II absorption as a guide for
setting limits (shown by vertical dashed lines) for integrating the
apparent optical depths $\tau_{\rm a}(v)$ for the Ni~II lines, as we
discuss in \S\protect\ref{genl}.  Without the aid of the S~II line, the
end points of the Ni~II features would otherwise be difficult to define
with much precision.  The dotted lines near the bottoms of the panels
show twice the expected rms amplitude of the noise in each pixel caused
by statistical fluctuations in the photon counts.  The value for
$\log\tau_{\rm a,int}$ for the Ni~II lines is about 1.1, which places
this pair of measurements just slightly below a median value of 1.2 for
all of the measurements shown in Fig.~\protect\ref{compare2_plot}. The
two features near the left edge of the lower panel are the 1316.5 and
1316.6$\,$\AA\ lines of S~I.\label{vstack_plot}}
\end{figure}

For our study of the absorption pairs at 1317 and 1741$\,$\AA, we used
spectroscopic data for 21 stars gathered from our earlier study of C~I
absorption features  (Jenkins \& Tripp 2001) that employed the E140H and
E230H echelle gratings of the Space Telescope Imaging Spectrograph
(STIS) on {\it HST\/} (observing program nrs.~8043 and 8484).  The
design and performance of STIS have been discussed by Woodgate et al. 
(1998)  and Kimble et al.  (1998).  The spectra had resolving powers
$\lambda/\Delta\lambda = 200,000$ because they were recorded with a very
narrow entrance slit on the spectrograph ($0\farcs 1\times 0\farcs03$),
and the detector was read out with twice the normal spatial sampling
rate.  We described in our previous article  (Jenkins \& Tripp 2001) the
special data reduction techniques that we used to preserve the high
spectral resolution without showing a high-frequency signal caused by a
sensitivity imbalance of adjacent Hi-Res pixels of the STIS MAMA
detector.  For the 1741$\,$\AA\ line, we could combine the recordings in
two adjacent echelle orders to increase the signal-to-noise ratio.  To
show the quality of the spectra, we present in Figure~\ref{vstack_plot}
some examples of the E140H and E230H data that we used.

%\placefigure{vstack_plot}

The observations by Jenkins \& Tripp  (2001) did not cover the Ni~II
line at 1370$\,$\AA, so an entirely separate collection of observations
was used to compare the 1317$\,$\AA\ transition to the one at
1370$\,$\AA.  Most of the data were taken from SNAP observations of many
stars in the Milky Way (program IDs 8241, 8662, 9434 initiated by
J.~Lauroesch) that are held in the MAST archive at the Space Telescope
Science Institute, with a few additional targets from the observations
taken for program nrs. 7137, 7270, 7301, and 8487.  The spectra were
recorded for 48 different stars using the E140H and E140M gratings with
wavelength resolving powers $\lambda/\Delta\lambda = 45,800$ and
110,000, respectively  (Kim Quijano et al. 2003).  A subset of these
stars was used for a validation exercise with another Ni~II line at
1454$\,$\AA\ discussed in \S\ref{f(1454)}.

For all of the spectra in both studies, the Ni~II features were always
visible.  They had equivalent widths that ranged from 8 to 140$\,$m\AA. 
The signal-to-noise ratios ($S/N$ per pixel)\footnote{Multiply by
$\sqrt{2}$ to obtain the $S/N$ per resolution element.} ranged from 18
to 50 (Fig.~\ref{vstack_plot} shows spectra with $S/N\approx 30$).  For
the star with the weakest lines, the equivalent widths of the Ni~II
lines were at least 3 times their respective $1\sigma$ uncertainties. 
We restricted our investigation to stars with large projected rotation
velocities, $v\sin i$, in order to avoid ambiguous continuum placement
problems and blending with stellar lines.  In a few cases, velocity
components that were well separated from each other could be measured
separately.

\section{Analysis}\label{analysis}
\subsection{General Considerations}\label{genl}

We fitted continua to the spectra using least-squares fits to Legendre
polynomials in places that were free of spectral features,\footnote{We
used care to avoid wavelengths near the 1316.5 and 1316.6$\,$\AA\
features of S~I.  Often these lines were not visible or barely visible;
the S~I lines that appear in the bottom panel of Fig.~\ref{vstack_plot}
are unusually strong.} following the methods outlined by Sembach \&
Savage  (1992).  Errors in line strength caused by continuum
uncertainties can be evaluated from the error matrix for the polynomial
coefficients, but we have found from past experience that it is wise to
multiply these uncertainties by 2 to allow for certain arbitrary
assumptions, such as the choice of wavelengths over which the
intensities are used to define the continuum.  Another source of error
is the uncertainty of photon counts within the absorption line, as
indicated by the STIS intensity error vector.  This error should be
independent of the one related to the continuum uncertainty.  Hence, we
could add the two in quadrature to evaluate the overall uncertainty in
the measurement.  For one of our line-pair comparisons (1317 vs.
1370$\,$\AA) we found that the magnitudes of the apparent random
deviations away from a simple relationship between the overall line
strengths was slightly larger than expected, indicating that additional
small errors are probably present.  (We will return to this issue
later.)   A reasonable interpretation for these additional errors is
that they arise either from variations in the detector sensitivity
(i.e., ``fixed-pattern noise'') that were not fully corrected, or from
inaccurate scattered light corrections.

For weak lines in the presence of noise, it is important to impose rigid
constraints on the wavelength limits for the measurement.  Otherwise, we
may introduce errors in the comparison if the limits changed from one
line to the next, and moreover, systematic biases in the measurements
can arise if the lines' own appearances are used as guides for the end
points  (Joseph 1989).  Thus, we defined the velocity limits for
interstellar material along each line of sight using the S~II line at
1250.578$\,$\AA.  This line is usually much stronger than the Ni~II
lines, so that it is easier to see the full extent of the absorptions
above the noise, but not so strong that small, irrelevant wisps of high
velocity material cause an unnecessary consideration of the more extreme
velocities.  Another advantage in using a strong line to define the
velocity limits is that we can be more certain that the continuum
definition occurs at wavelengths that are fully outside the region where
any absorption might occur.  The principle invoked here for defining the
velocity limits from the S~II feature is illustrated in
Fig.~\ref{vstack_plot}.

To simplify our comparison of line strengths, we compute the apparent
optical depths $\tau_a(v)=\ln [I_{\rm cont}(v)/I(v)]$ over all relevant
velocities $v$ within the absorption feature  (Savage \& Sembach 1991). 
The ratios of these quantities at any velocity for two lines should
indicate the relative values of $f\lambda$, provided there are no
narrow, unresolved structures that are saturated within the profiles 
(Savage \& Sembach 1991; Jenkins 1996).  The largest value of
$\tau_a(v)$ that we ever found was 1.4, which means that at no point in
any of the spectra were the lines heavily saturated.

\subsection{The $f$-value of the 1317$\,$\AA\ Line}\label{f(1317)}

\begin{figure}
\epsscale{1.0}
\plotone{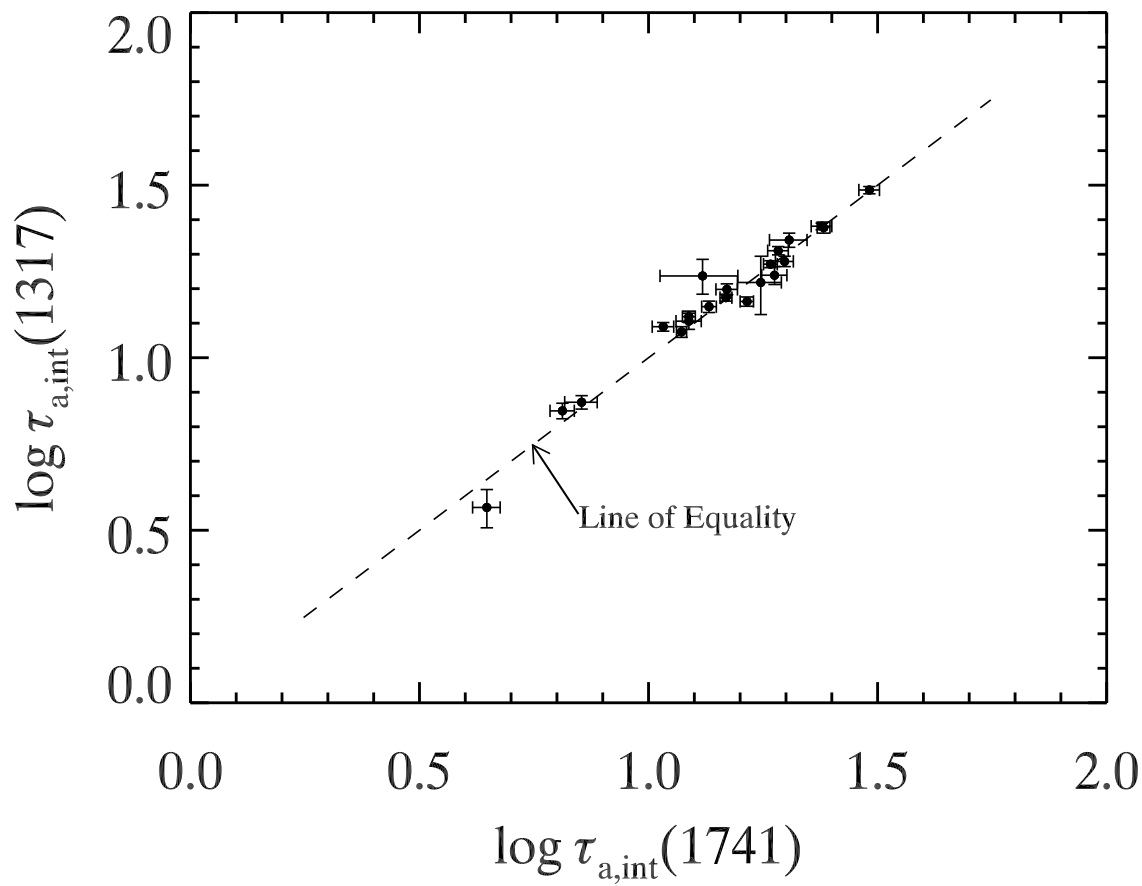}
\caption{A comparison of the logarithms of the integrated apparent
optical depths $\int \tau_a(v) dv$ (in ${\rm km~s}^{-1}$) for the
1317$\,$\AA\ and 1741$\,$\AA\ lines for 21 different stars.  The trend
line showing equal line strengths is shown by a dashed line; this line
is very nearly equal to our best-fit determination for the difference in
logarithmic strengths of the lines, $R=+0.005\,{\rm dex}$ that was
derived from a weighted mean using Eqs.~\protect\ref{R} and
\protect\ref{W}.\label{compare2_plot}}
\end{figure}

Figure~\ref{compare2_plot} shows comparisons in different spectra of the
apparent optical depths for the 1317 and 1741$\,$\AA\ lines, integrated
over velocity, i.e., $\tau_{\rm a,int}=\int \tau_{\rm a}(v)dv$.  It is
clear from this figure that the 1317$\,$\AA\ line has about the same
strength as the one at 1741$\,$\AA.  From the fact that all measurements
fall very near a single diagonal line with a unit slope in this diagram,
it is evident that stars with strong Ni~II lines do not show a ratio of
line strengths that is appreciably different from cases showing much
weaker features.  This increases our confidence that we are not being
misled by unrecognized saturation effects.

%\placefigure{compare2_plot}

Our best determination for $R$, a quantity that we designate as the
logarithm of the value of $f\lambda$ for the 1317$\,$\AA\ transition
divided by that for the 1741$\,$\AA\ one, was derived from a weighted
mean of the individual differences in the logarithms of the integrated
apparent optical depths,
\begin{equation}\label{R}
R=\sum \left[\log \tau_{\rm a,int}(1317)-\log \tau_{\rm
a,int}(1741)\right]_i\,W_i\bigg/\sum W_i
\end{equation}
where each weight $W_i$ is given by the inverse square of the error of
each comparison,
\begin{equation}\label{W}
W_i=\Big\{\sigma\big[\log \tau_{\rm a,int}(1317)\big]^2+\sigma\big[\log
\tau_{\rm a,int}(1741)\big]^2\Big\}^{-1}~,
\end{equation}
with $\sigma\big[\log \tau_{\rm a,int}(1317~{\rm or}~1741)\big]$ being
the uncertainty of the respective $\log \tau_{\rm a,int}$ value arising
from the combination of continuum and photon counting errors.  With the
use of Eqs.~\ref{R} and \ref{W} we found that
\begin{equation}\label{result2}
R=+0.005\pm 0.006\,{\rm dex~.}
\end{equation}
Our determination $\chi^2=22.1$ for 20 degrees of freedom seems
reasonable, indicating that our formally derived errors are probably
correct.  The formal uncertainty for $R$ given in Eq.~\ref{result2} was
derived from the expression $\sigma(R)=\left(\sum W_i\right)^{-0.5}$.

To determine the $f$-value of the 1317$\,$\AA\ transition, we take the
product of $f(1741)=0.0427$ measured by Fedchak et al.  (2000) and
listed by Morton  (2003), $10^R$, and the ratio of wavelengths
(1741/1317) to obtain $f(1317)=0.0571$.  Our value for $\sigma(R)$ is
much smaller than the probable uncertainty of 0.04~dex estimated by
Fedchak et al.  (2000) for the value of $\log f(1741)$, which served as
our comparison standard.  This error thus dominates the relative
uncertainty of the $f$-value of the 1317$\,$\AA\ line.

\subsection{The $f$-value of the 1370$\,$\AA\ Line}\label{f(1370)}

\begin{figure}
\plotone{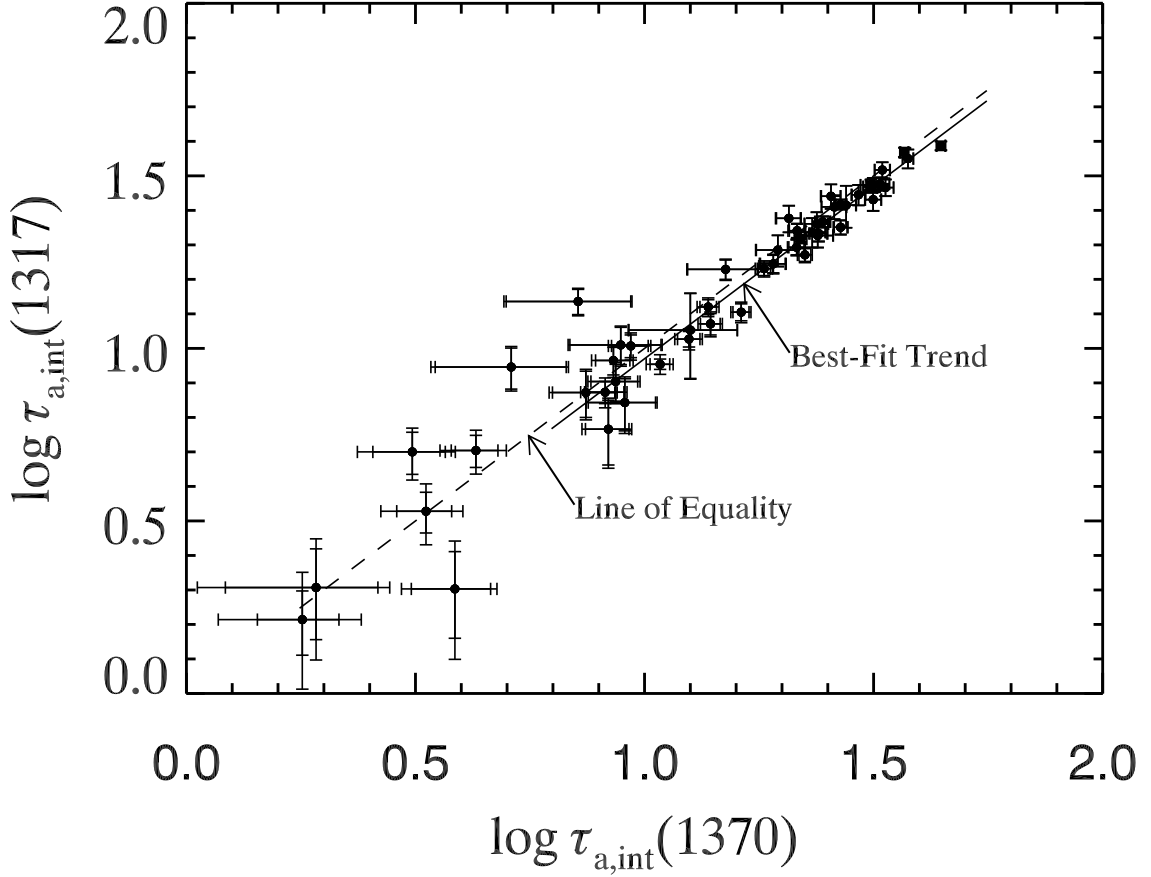}
\caption{A comparison of the logarithms of the integrated apparent
optical depths $\int \tau_a(v) dv$ for the 1317$\,$\AA\ and 1370$\,$\AA\
lines for 48 different stars.  The inner ticks on the error bars
indicate the originally computed uncertainties, while the outer ones
show the artificially increased errors needed to make the $\chi^2$
values roughly equal to the number of degrees of freedom.  Two trend
lines are shown: the dashed line shows the locus of equal line
strengths, while the solid line indicates where the transitions have a
logarithmic difference of strengths $R=-0.030\,{\rm dex}$.  As with the
data shown in Fig.~\ref{compare2_plot}, Eqs.~\protect\ref{R} and
\protect\ref{W} were used to evaluate $R$, but with the designations
``1741'' being replaced by ``1370.''\label{compare_plot}}
\end{figure}

Now that we have derived our value for $f(1317)$, we can use our other
set of observations that had both the 1317 and 1370$\,$\AA\ features to
derive $f(1370)$.  Figure~\ref{compare_plot} shows our comparison of
these two lines, shown in the same style as that of
Fig.~\ref{compare2_plot}.  Again we use Eqs.~\ref{R} and \ref{W} (but
this time replacing the 1741$\,$\AA\ line with the 1370$\,$\AA\ one) to
derive $R=-0.031\,$dex with $\sigma(R)=0.004\,$dex.

%\placefigure{compare_plot}

Unlike what we had found for our study of the 1317/1741 line pairs, we
obtained indications that our calculated line measurement errors
slightly underestimated the true ones, since we found that $\chi^2=70.1$
for 51 degrees of freedom.\footnote{While only 48 stars were measured, 4
of them had extra, well separated velocity components that could be
measured independently.}  To properly acknowledge that the scatter of
the results exceeded our initial expectations, we created an additional,
ad-hoc error term $\sigma({\rm extra})=0.5\,{\rm km~s}^{-1}$ that we
felt should be factored into our determinations of $\tau_{\rm a,int}$. 
The square of this term was added to the sums of the squares of the
errors attributable to just the deviations caused by continuum and noise
uncertainties, so that the value of $\chi^2$ was ultimately reduced to
about 50.  With these slightly larger errors, the re-evaluation of
Eqs.~\ref{R} and \ref{W} gave our final result,
\begin{equation}\label{result1}
R=-0.030\pm 0.006\,{\rm dex~.}
\end{equation}

In a calculation similar to the one for the 1317$\,$\AA\ line, we now
evaluate the product of $f(1317)=0.0571$ that we found earlier
(\S\ref{f(1317)}), $10^{-R}$, and the wavelength ratio to obtain
$f(1370)=0.0588$.  By combining the errors in the determinations of the
two values of $R$ that must be used (the ones in Eqs.~\ref{result2} and
\ref{result1}) to link ultimately to the $f$-value of the 1741$\,$\AA\
line, we obtain a formal relative uncertainty of 0.009$\,$dex, which
once again is small compared to the 10\% error in the comparison line.

\subsection{Validation of our Method using the 1454$\,$\AA\
Line}\label{f(1454)}

The strength of the Ni~II transition to the $3d^8(^1D)4p~^2D^{\rm
\,o}_{5/2}$ level at 1454.842$\,$\AA\ has been determined by Fedchak et
al.  (2000), although only to an accuracy of 25\% (0.12$\,$dex).  As a
demonstration, we can repeat the analysis that we performed in
\S\ref{f(1370)} for this line, so that we can gain some assurance that
our comparison method is sound.  In fact, this test is more demanding
than the actual studies that we carried out, since the strength of the
1454$\,$\AA\ is substantially less than those of the other three lines.

\begin{figure}
\plotone{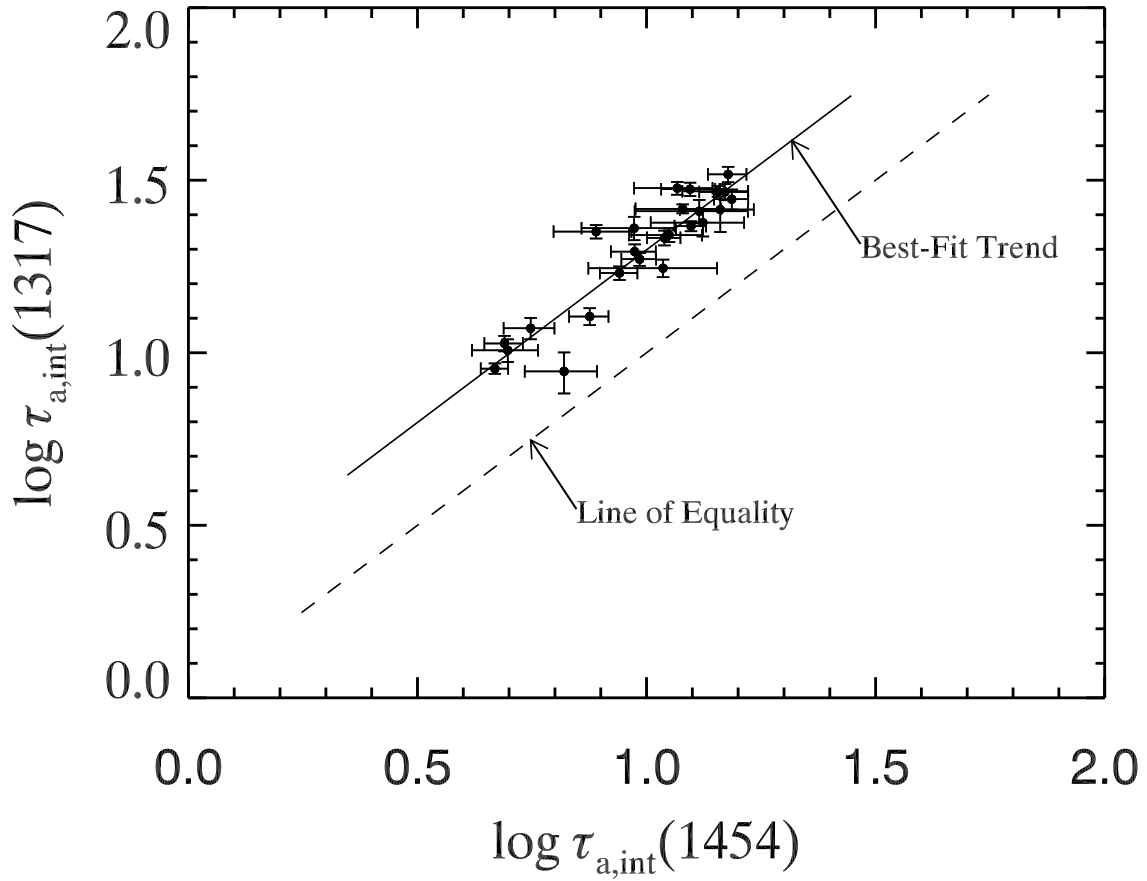}
\caption{A validation of our method, where we compare the logarithms of
the integrated apparent optical depths $\int \tau_a(v) dv$ (in ${\rm
km~s}^{-1}$) for the 1317$\,$\AA\ and 1454$\,$\AA\ lines for 25
different stars.  Repeating the analysis that we had done earlier for
deriving the other two $f$-values gave $f(1454)=0.0260\pm 0.0026$, which
is somewhat less than the value $0.0323\pm 0.008$ given by Fedchak et
al.  (2000), but still within their stated 25\%
error.}\label{compare3_plot}
\end{figure}

Figure~\ref{compare3_plot} shows our comparison of the 1317$\,$\AA\ and
1454$\,$\AA\ lines, shown in the same style as in
Figs.~\ref{compare2_plot} and \ref{compare_plot}.  For these data, once
again we used Eqs.~\ref{R} and \ref{W} (replacing the 1741$\,$\AA\ line
with the 1454$\,$\AA\ line) to obtain
\begin{equation}\label{result3}
R=0.298\pm 0.012\,{\rm dex~.}
\end{equation}
which yielded $\chi^2=16.7$ for 23 degrees of freedom.  From the above
value of $R$, we infer from a repeat of the calculations discussed
previously that $f(1454)=0.0260\pm 0.0026$.  This result is consitent
with the value $0.0323\pm 0.008$ derived by Fedchak et al.  (2000) to
within their stated error.

\section{Summary}\label{summary}

In acknowledging the importance of deriving gas-phase abundances of Ni
in various astrophysical systems, we have recognized a need for deriving
an $f$-value for the Ni~II transition at 1317$\,$\AA, as well as the
desirability of obtaining an improved accuracy for the $f$-value of the
line at 1370$\,$\AA.  For the first of these two goals, we have compared
the integrated apparent optical depths $\tau_{\rm a,int}=\int \tau_{\rm
a}(v)dv$ of interstellar Ni~II absorption features at 1317$\,$\AA\ to
those of their counterparts 1741$\,$\AA\ seen in high-resolution spectra
of 21 stars recorded by STIS on {\it HST}.  Since the $f$-value of the
1741$\,$\AA\ line has been determined by laboratory measurements to a
good accuracy (10\%), it can serve as a comparison standard for the
1317$\,$\AA\ line.  We have deliberately avoided trying to compare the
1317$\,$\AA\ transition to other, weaker ones, so that we can bypass the
uncertainties that are created by possible unresolved, saturated
substructures within the line profiles.  Nevertheless, the success of
our demonstration given in \S\ref{f(1454)} shows that our method works
satisfactorily even when there is a moderately large disparity of line
strengths.  We used the strong profiles of S~II absorptions at
1250$\,$\AA\ to define the velocity end points of the Ni~II profiles,
which on some occasions were difficult to measure.

Our analysis indicated that $f(1317)=0.0571\pm 0.006$, where the error
is dominated by the uncertainty in the $f$-value of the comparison line. 
It is important to note that this value is only 0.39 times as large as
the approximate theoretical value supplied by Kurucz that was listed in
Morton's  (1991) earlier compilation (and does not appear in the most
recent listing of 2003).  Following our derivation of $f(1317)$, we
examined spectra of 48 stars (different from the ones used for the
1317/1741 comparison) that had both the 1317 and 1370$\,$\AA\ lines, so
that we could improve upon the previously published $f$-value of the
1370$\,$\AA\ line using our value of $f(1317)$ as a comparison standard. 
This analysis yielded the result $f(1370)=0.0588\pm 0.006$, which is
0.76 times the revised value of Zsarg\'o \& Federman  (1998)  given in
Morton's  (2003)  compilation of $f$-values (and 0.45 times the value
given in his earlier listing in 1991, again based on the theoretical
values of Kurucz).  Our new result for $f(1370)$ differs from the
rescaled value of Zsarg\'o \& Federman  (1998) by less than their
declared uncertainty of 30\%.  The transition probabilities listed by
Fritzsche et al.  (2000) (from their length calculations using
experimentally determinied energies) $A_{21}(1317)=1.95\times 10^8\,{\rm
s}^{-1}$ and $A_{21}(1370)=3.55\times 10^8\,{\rm s}^{-1}$ are equivalent
to $f(1317)=0.0676$ and $f(1370)=0.0666$, which are respectively only
18\% and 13\% higher than our determinations.

\acknowledgements

We thank David~Bowen and Don~Morton for their comments on early drafts
of this paper.  This research was supported by NASA HST archival grant
HST-AR-09534.01-A from the Space Telescope Science Institute, which is
operated by the Association of Universities for Research in Astronomy,
Incorporated, under NASA contract NAS5-26555.

\clearpage
\end{document}